\documentclass[a4paper,11pt]{article}
\usepackage[latin1]{inputenc}
\usepackage[round]{natbib}
\usepackage{color} 
\usepackage[mathcal]{euscript}
\usepackage{graphicx}
\usepackage{subcaption}
\usepackage{amsfonts}
\usepackage{amssymb}
\usepackage{geometry}
\usepackage{longtable}
\usepackage{fancyhdr}
\usepackage{indentfirst}
\usepackage{lscape}
\usepackage{float}
\usepackage{url}
\usepackage{bbm}
\usepackage{amsmath}

\usepackage{algorithm}
\usepackage{algpseudocode}
\usepackage{pifont}
\usepackage{setspace}

\DeclareMathOperator*{\argmin}{argmin}   % Jan Hlavacek
\DeclareMathOperator{\tr}{tr}

\DeclareMathOperator*{\diag}{diag}

\usepackage[colorlinks=true]{hyperref}
\hypersetup{
	allcolors = {blue},
}
\makeatletter
\begin{document}

\title{\textbf{Decoupling Shrinkage and Selection\\ in Gaussian Linear Factor Analysis}}
\author{Henrique Bolfarine \thanks{Institute of Mathematics and Statistics, University of S\~ao Paulo, S\~ao Paulo, Brazil, e-mail:bolfarin@ime.usp.br.} \and Carlos M. Carvalho\thanks{University of Texas McCombs School of Business, 2110 Speedway, Austin, Texas 78705, e-mail:carlos.carvalho@mccombs.utexas.edu.}\and  Hedibert F. Lopes\thanks{INSPER Institute of Education and Research, S\~ao Paulo, Brazil, e-mail:hedibertFL@insper.edu.br.}\and Jared S. Murray\thanks{University of Texas McCombs School of Business, 2110 Speedway, Austin, Texas 78705, e-mail:jared.murray@mccombs.utexas.edu.}}
\date{}

\maketitle
\begin{abstract}
\noindent Factor analysis is a popular method for modeling dependence in multivariate data. However, determining the number of factors and obtaining a sparse orientation of the loadings are still major challenges. In this paper, we propose a decision-theoretic approach that brings to light the relation between model fit,  factor dimension, and sparse loadings. This relation is done through a summary of the information contained in the multivariate posterior. To construct this summary we introduce a two-step approach. First, given posterior samples from the Bayesian factor analysis model, a series of point estimates with a decreasing number of factors, and different levels of sparsity is recovered by minimizing an expected penalized loss function. Second, the degradation in model fit between the posterior and the recovered estimates is displayed in a summary. In this step, a criterion is proposed for selecting the factor model with the best trade-off between fit, sparseness, and factor dimension. The findings are illustrated through a simulation study and an application to personality data. We use different prior choices to show the flexibility of the proposed method. 

\end{abstract}


\vspace{0.4cm}

\newpage

\tableofcontents

\newpage
\section{Introduction}\label{ref_INT}

Factor analysis is an important tool for modeling the dependence structure among variables. Over the years, factor analysis and related factor models have found their way into applications in different fields, such as economics, finance, and genomics \cite[see][and references therein]{fruehwirth2018sparse}. As
is well known, the main challenges surrounding the factor analysis model are the specification of the number of factors and obtaining a sparse and interpretable loadings matrix \citep{rovckova2016fast}. In this paper, we address these problems by proposing a novel decision-theoretic approach \citep[][]{bernardo2009bayesian}, that reveals the relation between model fit, factor dimension and a sparse representation of the loadings matrix. Our goal is not only obtaining interpretable point estimates for the factor loadings and uniqueness yielded from the decision analysis, but a summary of the information encapsulated in the multivariate posterior. To obtain the point estimates and perform the summary, we follow the approach presented by \cite{hanhcarvalho2015}, in which the decoupling shrinkage and selection (DSS) method is introduced in the context of regression models.  The essential element of the DSS approach is a selection strategy based on the posterior predictive distribution, which provides a relevant scale on which to consider whether a sparse lower dimensional version of the model has an adequate fit. Recently, these technique have been used for obtaining informative summaries in seemingly unrelated regressions \citep{puelz2017variable}, graphical models \citep{bashir2018post}, functional regressions \citep{kowal2020bayesian1}, nonlinear regressions \citep{woody2021model}, and in time-varying parameter models \citep{Huber2021}.

The general framework of the DSS method for factor analysis (DSSFA) can be summarized in two steps. First, provided that posterior samples of the Bayesian factor analysis model are available, a series of optimal point estimates indexed by a decreasing number of factor dimensions, and  with different levels of sparsity in the loadings is obtained through the minimization of an expected penalized loss function. Second, we generate a posterior summary that encapsulates the loss in fit between the full factor model, produced by the posterior distribution, and the model generated by sparse lower dimension estimates, referred to as a simple model. This summary is reported in a plot that can be visually inspected in search of the model that yields the best fit. We also propose a criterion, that selects the factor model with the best trade-off between fit, sparseness, and factor dimension in an automated fashion.

As we will see in the next section, the DSSFA approach connects different strands of the factor analysis literature by incorporating ideas from parametric methods, where posterior samples are obtained via well established stochastic algorithms \citep{lopes2004bayesian,bernardo2003bayesian,carvalho2008high,fruehwirth2018sparse}, to recent methods that do not impose identifying restrictions on its inference algorithm, and do not require pre-specification of the factor dimension \citep{bhattacharya2011sparse,legramanti2020bayesian}. Additionally, unlike hard thresholding rules and classical methods based on information criteria \citep{scwarz1978,akaike1987factor}, our method allows for simultaneous model and factor loadings selection. To the best of our knowledge, this is the first paper to present model selection for factor analysis as a decision problem, and the first to extend DSS procedure to latent variable modeling. 

This paper is organized as follows. The remainder of this section reviews the Bayesian factor analysis model. Section \ref{sec3} introduces the framework of the proposed approach. In Section \ref{sim.study}, we follow the simulation design of \cite{man2020mode}, where we compare our proposed method with the marginal likelihood estimate for selecting the number of factors \citep{lopes2004bayesian,newton1994approximate}, and Bayes model averaging (BMA) \citep{hoeting1999bayesian} for covariance matrix estimation. In Section \ref{data.ana}, we apply DSSFA in a subset of the big five personality traits data, and evaluate how our method interacts with over-fitted priors \citep{bhattacharya2011sparse,legramanti2020bayesian}, resulting in useful posterior summaries and interpretable factor loadings. Finally, some conclusions are given in Section \ref{concl}. The methods and data presented here are available at \texttt{https://github.com/hbolfarine/dssfa}.

\subsection{Notations  for the basic factor analysis model}\label{sec11}

In the basic factor analysis model,  $\boldsymbol{y}_i = (\boldsymbol{y}_{1i},\dots,\boldsymbol{y}_{pi})^T$ is a $p$-dimensional vector of observations in a random sample $\boldsymbol{y} = (\boldsymbol{y}_1,\dots,\boldsymbol{y}_n)^T$, which relates to a $k$-dimensional vector, with $k\leq p$, of common latent factors $\boldsymbol{f}_i$ through
\begin{equation}\label{eq2}
\boldsymbol{y}_i =\boldsymbol{B}\boldsymbol{f}_i+\boldsymbol{\epsilon}_i,
\end{equation}
where $\boldsymbol{B}$ is a $p\times k$  factor loadings matrix,  $\boldsymbol{f}_i$ is normally distributed as $N_k(\boldsymbol{0},\boldsymbol{I}_k)$, and $\boldsymbol{\epsilon}_i$ is the  idiosyncratic error vector with dimension $p$. In model \eqref{eq2}, we assume: (i) $\boldsymbol{\epsilon}_i \sim N_p(\boldsymbol{0},\boldsymbol{\Sigma})$, with $\boldsymbol{\Sigma} = \diag(\sigma_1^2,\dots,\sigma_p^2)$ and  $\sigma_j^2>0$, for all $j = 1,\dots,p$, and (ii) $\boldsymbol{f}_r$ and $\boldsymbol{\epsilon}_t$ are independent for all $r$ and $t$. These assumptions and the model in \eqref{eq2} imply that the marginal distribution of $\boldsymbol{y}_i$ in relation to $\boldsymbol{f}_i$ is given by $N_{p}(\boldsymbol{0},\boldsymbol{\Omega})$, where  $\boldsymbol{\Omega}= \boldsymbol{B}\boldsymbol{B}^T+ \boldsymbol{\Sigma}$ is the covariance matrix.  

To complete the Bayesian specification of model \eqref{eq2}, let $p(\boldsymbol{B},\boldsymbol{\Sigma}) = p(\boldsymbol{B})p(\boldsymbol{\Sigma})$ be the prior choice on the loadings matrix and uniqueness. The common choice for $p(\boldsymbol{B})$ specifies truncated normal priors for the diagonal elements, and normal priors for the remaining lower triangular elements of $\boldsymbol{B}$. This configuration is known as positive lower triangular constrain (PLT). For the elements of the uniqueness $\boldsymbol{\Sigma}$, inverse-gamma priors are preferred. These choices are convenient, because they lead to conjugate forms that are straightforward to compute using Gibbs sampler \citep{gamerman2006markov}. The goal of the PLT constrain is to address the problem of posterior identifiability of $\boldsymbol{B}$, in which one can obtain the same covariance matrix $\boldsymbol{\Omega}$, defined by \eqref{eq2}, by multiplying $\boldsymbol{B}$ by an orthonormal matrix $\boldsymbol{P}$ where $\boldsymbol{P}\boldsymbol{P}^T = \boldsymbol{I}_k$ \citep[see][]{geweke1996measuring,lopes2004bayesian}. An analysis of recent approaches to prior choices on the problem of posterior identifiability can be seen in \cite{man2020mode}.

Importantly, under our approach, no modeling assumptions are made regarding prior choices provided samples from the posterior marginals distributions, $p(\boldsymbol{B}|\boldsymbol{y})$, and $p(\boldsymbol{\Sigma}|\boldsymbol{y})$ are available. As we will see in the next section, we only need the information of the factor dimension of the posterior loadings, $k$. Thereafter, throughout this paper, we assume the posterior on $\boldsymbol{B}$ has $p\times k$ dimension, and is referred to as the full model posterior.

\section{Decoupling Shrinkage and Selection in Factor Analysis}\label{sec3}

In Subsection \ref{expec_dssfa}, we establish a decision analysis framework for obtaining estimates for the factor analysis model.  In Subsection \ref{gsum}, we present the posterior summary and a criterion to perform model selection. We finish with an overview of our approach, and a toy example in Subsection \ref{resum}.

\subsection{DSSFA estimates}\label{expec_dssfa}

Here, we define the recovery of point estimates for the factor analysis model as a decision problem \citep{bernardo2009bayesian,berger2013statistical}. Under the proposed framework, the full model posterior of the Bayesian factor analysis model, presented in Subsection \ref{sec11}, is fundamental to determine optimal point estimates that preserve the fitness of the full model. Commonly, in the DSS framework, the posterior predictive distribution is used to obtain the best accuracy in prediction given future observations \citep{hanhcarvalho2015,woody2021model,kowal2021fast}. However, in this paper, we have a special interest in factor analysis model fit, and thus we focus solely on the distribution of the posterior parameters rather than the predictive distribution without loss in the results.

We start by selecting an appropriate loss function. Since the factor analysis model defined in Subsection \ref{sec11} depends directly on the covariance matrix, a natural choice for evaluating the model fit is the negative log-likelihood of the multivariate normal distribution. This loss is strongly associated with Stein's loss and the Kullback-Leibler divergence between two normal distributions. We further expand the proposed loss, by incorporating a complexity penalty that embodies the trade-off between the fit and model simplicity resulting in
\begin{equation}\label{eq_loss}
{\cal L}_{\lambda}(\boldsymbol{\Omega},\boldsymbol{\tilde{\Omega}}) = \log|\boldsymbol{\tilde{\Omega}}|+\tr\left(\boldsymbol{\tilde{\Omega}}^{-1}\boldsymbol{\Omega}\right)+\lambda P(\boldsymbol{\tilde{\Omega}}),
\end{equation}
where $\boldsymbol{\tilde{\Omega}}$ is a $p\times p$ positive semi-definite symmetric matrix, $\boldsymbol{\Omega}$ is a $p\times p$ covariance matrix defined the assumptions presented in Subsection \ref{sec11}, $P(\cdot)$ is the complexity penalty, and $\tr(A)$ is the trace of matrix $A$. The complexity parameter $\lambda\geq 0$ controls the trade-off between model fit and parsimony resulting in a simple model. It is important to reiterate the distinction between $\boldsymbol{\tilde{\Omega}}$ and the covariance matrix $\boldsymbol{\Omega}$ in \eqref{eq_loss}. As seen in Subsection \ref{sec11}, we have $\boldsymbol{\Omega} = \boldsymbol{B}\boldsymbol{B}^T+\boldsymbol{\Sigma}$, where  $\boldsymbol{B}$ and  $\boldsymbol{\Sigma}$ are parameters of the Bayesian factor analysis model and thus are associated with the prior distribution $p(\boldsymbol{B},\boldsymbol{\Sigma})$. By comparison, is not coherent to place a prior on $\boldsymbol{\tilde{\Omega}}$ and $\lambda$ as they define an action and a penalty, respectively, in the proposed decision analysis framework.

Next, we proceed with the decision analysis and obtain optimal point estimates. Hence, given a complexity level $\lambda$, by minimizing the posterior expectation of \eqref{eq_loss}, over the posterior distribution of the factor analysis model presented in Subsection \ref{sec11}, results in
\begin{equation}\label{eq_3_2}
\boldsymbol{\hat{\Omega}}_{\lambda}  \equiv \argmin_{\boldsymbol{\tilde{\Omega}}} E_{\boldsymbol{\theta}|\boldsymbol{y}}\left[{\cal L}_{\lambda}(\boldsymbol{\Omega},\boldsymbol{\tilde{\Omega}})\right],
\end{equation}
where $\boldsymbol{\hat{\Omega}}_{\lambda}$ is the optimal point estimate, and $\boldsymbol{\theta} = (\boldsymbol{B},\boldsymbol{\Sigma})$. We simplify the expectation in expression \eqref{eq_3_2}, as follows
\begin{eqnarray}\label{loss_eq}
E_{\boldsymbol{\theta}|\boldsymbol{y}}\left[{\cal L}_{\lambda}(\boldsymbol{\Omega},\boldsymbol{\tilde{\Omega}})\right] &=&E_{\boldsymbol{\theta}|\boldsymbol{y}}\left[\log|\boldsymbol{\tilde{\Omega}}|+\tr\left(\boldsymbol{\tilde{\Omega}}^{-1}\boldsymbol{\Omega}\right)\right] \nonumber+\lambda P(\boldsymbol{\tilde{\Omega}})\\
&=& \log|\boldsymbol{\tilde{\Omega}}|+\tr\left(\boldsymbol{\tilde{\Omega}}^{-1}\overline{\boldsymbol{\Omega}}\right)+\lambda P(\boldsymbol{\tilde{\Omega}}),
\end{eqnarray}
where $ \overline{\boldsymbol{\Omega}} =E_{\boldsymbol{\theta}|\boldsymbol{y}}[\boldsymbol{\Omega}]$ is the posterior mean of the covariance matrix. By integrating over the marginal posterior  distributions of the Bayesian factor  analysis model, $p(\boldsymbol{B}|\boldsymbol{y})$, and $p(\boldsymbol{\Sigma}|\boldsymbol{y})$, we  have $\overline{\boldsymbol{\Omega}} = \overline{\boldsymbol{BB}^T}+\overline{\boldsymbol{\Sigma}}$, where $\overline{\boldsymbol{BB}^T} = E_{\boldsymbol{B}|\boldsymbol{y}}\left[\boldsymbol{B}\boldsymbol{B}^T\right]$, and $\overline{\boldsymbol{\Sigma}} = E_{\boldsymbol{\Sigma}|\boldsymbol{y}}\left[\boldsymbol{\Sigma}\right]$ are the expected posterior values.

We make some observations on expression \eqref{loss_eq}: (i) the resulted expected loss function depends uniquely on the posterior mean of the covariance matrix $ \overline{\boldsymbol{\Omega}}$, thus it is agnostic to prior choice, (ii) $\overline{\boldsymbol{\Omega}}$ is robust to factor rotation, and therefore the identifiability of the factor analysis model on prior choice is not a concern. Furthermore, \eqref{loss_eq} is valuable for obtaining covariance matrix point estimates. For example, if we assume the unpenalized case, $\lambda = 0$, the solution of \eqref{eq_3_2} is $\boldsymbol{\hat{\Omega}}_{\lambda = 0} = \boldsymbol{\overline{\Omega}}$, the posterior covariance matrix. 

However, our interest lies in optimal decisions for the factor analysis model, with a special interest in the structure of the loadings matrix $\boldsymbol{B}$. Therefore, in our decision framework, we assume $\boldsymbol{\tilde{\Omega}} = \boldsymbol{\tilde{B}}\boldsymbol{\tilde{B}}^T+\boldsymbol{\tilde{\Sigma}}$, where ${\boldsymbol{\tilde{B}}}$ is a $p\times \tilde{k}$ matrix, $\boldsymbol{\tilde{\Sigma}}$ is a $p\times p$ diagonal matrix, with positive entries, and $\tilde{k} \in \{1,2,\dots\}$, are elements of the decision analysis. Notably, $\tilde{k}$ is a choice of dimension summary on the resulting loadings estimates. For instance, for smaller values of $\tilde{k}$, the resulting optimal estimates have lower dimensions, and when $\tilde{k}=k$, results in an estimate with the dimension of the full model posterior. In this paper, no further restrictions are imposed on the identifiability of $\boldsymbol{\tilde{B}}$, although a $\tilde{k} \times \tilde{k}$ symmetric matrix $\boldsymbol{\tilde{\Phi}}$ could be introduced to the decomposition of  $\boldsymbol{\tilde{\Omega}} = \boldsymbol{\tilde{B}}\boldsymbol{\tilde{\Phi}}\boldsymbol{\tilde{B}}^T+\boldsymbol{\tilde{\Sigma}}$, resulting in a decision analysis for the oblique factor model \citep{thurstone1947multiple}.

We further reiterate that ${\boldsymbol{\tilde{B}}}$, $\boldsymbol{\tilde{\Sigma}}$ and $\tilde{k}$ are actions in the decision analysis, and thus are not subject to the prior specification of the Bayesian factor analysis model.  By comparison, a prior on $\boldsymbol{B}$ may indicate our preference for sparse loadings, although it does not guarantee sparsity in the posterior. Conversely, under our framework, we can expand the complexity penalty $P(\cdot)$ to $\boldsymbol{\tilde{B}}$, allowing for a sparse representation of the estimates regardless of prior choice.  Many choices of complexity penalty are available, but to shrink the factor loadings to zero, we consider sparsity-inducing penalties such as $\ell_1$-penalty, which are commonly used for model selection in regression settings \citep{tibshirani1996regression}. Thus, we update the complexity penalty as $P(\boldsymbol{\tilde{\Omega}}) = \|\boldsymbol{\tilde{B}}\|_1$, where $\|A\|_1 = \sum_{j= 1}^p\sum_{q = 1}^k|a_{jq}|$, for $a_{jq}\in A$.  Furthermore, the chosen penalty on $\boldsymbol{\tilde{B}}$ can be used to prevent the non-identifiability of the loadings matrix \citep{scharf2019should}.

Finally, we obtain optimal estimates $\boldsymbol{\hat{B}}_{\tilde{k},\lambda}$, and $\boldsymbol{\hat{\Sigma}}_{\tilde{k},\lambda}$, referred to as DSSFA estimates, for the factor loadings and uniqueness, respectively, and a given complexity level $\lambda$, by minimizing
\begin{equation}\label{eq_3_5}
(\boldsymbol{\hat{B}}_{\tilde{k},\lambda},\boldsymbol{\hat{\Sigma}}_{\tilde{k},\lambda})   \equiv \argmin_{\boldsymbol{\tilde{B}},\boldsymbol{\tilde{\Sigma}}}\left\{\log|\boldsymbol{\tilde{\Omega}}|+\tr\left(\boldsymbol{\tilde{\Omega}}^{-1}\overline{\boldsymbol{\Omega}}\right) + \lambda\|\boldsymbol{\tilde{B}}\|_1\right\},
\end{equation}
subject to $\boldsymbol{\tilde{\Omega}} = \boldsymbol{\tilde{B}}\boldsymbol{\tilde{B}}^T+\boldsymbol{\tilde{\Sigma}}$, and to different values of $\tilde{k}$. 

It is noteworthy that if there is an interest solely in the number of factors, the minimization in \eqref{eq_3_5}, can be performed with no penalty on the loadings  ($\lambda = 0$), for different values of $\tilde{k}$. In this paper, the maximum value for $\tilde{k}$ is chosen to be the dimension of the full model posterior, $k$, although the size of $\tilde{k}$ is only limited to the optimization procedure. 

An efficient algorithm exists for the chosen complexity penalty $P(\cdot)$. We use an off-the-shelf procedure implemented in the \texttt{fanc} package \citep{Hirose2016} from \texttt{R} \citep{citeR}.  See Supplementary Material \ref{A1.3} for the  \texttt{R} implementation of the optimization procedure. We recall that there are other possible methods to solve \eqref{eq_3_5} \cite[see][]{scharf2019should}.

\subsection{Posterior  summary}\label{gsum}

In this section, we propose a posterior summary that displays the trade-off between the change in factor dimension, sparse loadings  and the subsequent loss in model fit. The loss function in \eqref{eq_3} encapsulates information between fit and simplicity, then by replacing the action $\boldsymbol{\tilde{\Omega}}$, with the DSSFA covariance estimate, $\boldsymbol{\hat{\Omega}}_{\tilde{k},\lambda} = \boldsymbol{\hat{B}}_{\tilde{k},\lambda}\boldsymbol{\hat{B}}_{\tilde{k},\lambda}^{T}+\boldsymbol{\hat{\Sigma}}_{\tilde{k},\lambda}$, we can assess the impact of moving from the full model, with $\lambda = 0$, and $\tilde{k} = k$, to the sparse low-dimensional representation. This generates a sequence of loss functions, ${\cal L}_{\tilde{k},\lambda}(\boldsymbol{\Omega},\boldsymbol{\hat{\Omega}}_{\tilde{k},\lambda})$ (see Step 2, of Subsection \ref{resum}), that depend on the posterior distribution of the covariance matrix generated by the Bayesian factor model. This sequence is important to evaluate how the coefficients $\tilde{k}$ and $\lambda$, affect model fit in different dimension and levels of complexity. Thus, we summarize this change in a grid, which can be visually inspected in a plot, exposing the trade-off between fit, factor dimension, and sparseness. The models generated by the estimates $\boldsymbol{\hat{B}}_{\tilde{k},\lambda}$, produced by \eqref{eq_3_5}, whose columns were zeroed by the optimization procedure are discarded from the evaluation.

In the DSS framework, the scale of the distribution generated by the loss function of the full model posterior is relevant for judging acceptable fit \citep{hanhcarvalho2015,bashir2018post}. Hence, we consider models whose expected posterior loss, $E_{\boldsymbol{\theta}|\boldsymbol{y}}[{\cal L}_{\tilde{k},\lambda}(\boldsymbol{\Omega},\boldsymbol{\hat{\Omega}}_{\tilde{k},\lambda})]$, is within the quantile of the loss function of the full model. However, it is left to the end user to decide the best simple representation given the information displayed in the summary. 

In this paper we define a criterion that selects simultaneously the model with the lowest factor dimension, and the sparsest loadings matrix and that also preserves the fit of the full model. By considering the largest expected posterior loss, and the smallest factor dimension $\tilde{k}$ within the quantile of the loss function of generated by the full model, we have the maximum acceptable trade-off between the change in factor dimension, and sparse loadings. By adopting this criterion, we identify the simplest factor analysis model. Furthermore, this criterion allows for the factor model to be selected in an automated fashion, without the necessity of a visual inspection of the summary plot. In this paper, we considered quantiles between $95\%$ and $99\%$ of the loss function of the full model.  We also apply this criterion to a toy example in Subsection \ref{toyexemp}, a numerical example in Section \ref{sim.study}, and an empirical application in  Section \ref{data.ana}. 

\subsection{Method overview}\label{resum}

Before illustrating the proposed approach with a toy example, we present an overview of the DSSFA method summarized in two steps. We initiate our procedure by calculating the posterior covariance mean, $\boldsymbol{\overline{\Omega}} = \overline{\boldsymbol{B}\boldsymbol{B}^T}+\boldsymbol{\overline{\Sigma}}$, from the posterior distributions, $p(\boldsymbol{B}|\boldsymbol{y})$, and $p(\boldsymbol{\Sigma}|\boldsymbol{y)}$, where the number of factors was chosen as $k$. We approximate $\overline{\boldsymbol{BB}^T}\approx \frac{1}{M} \sum_{m = 1}^{M}\boldsymbol{B}_{(m)} \boldsymbol{B}_{(m)}^{T}$ and $\boldsymbol{\overline{\Sigma}} \approx  \frac{1}{M} \sum_{m = 1}^M\boldsymbol{\Sigma}_{(m)}$ where $\boldsymbol{B}_{(m)}$ and $\boldsymbol{\Sigma}_{(m)}$ are the posterior samples with $m = 1,2,\dots,M$. 

\begin{enumerate}
	\item[S.1] DSSFA estimates: Apply the  posterior mean of the covariance matrix $\overline{\boldsymbol{\Omega}}$, to the optimization procedure  in the package \texttt{fanc} to solve  \eqref{eq_3_5}. Obtain a sequence of sparse loadings and uniqueness, ($\boldsymbol{\hat{B}}_{\tilde{k},\lambda}$, $\boldsymbol{\hat{\Sigma}}_{\tilde{k},\lambda}$), for $\tilde{k} = 1, \dots,k$, indexed by  $\lambda  =  \lambda_0,\lambda_1,\dots,\lambda_l$, where $\lambda_{0} = 0$, and $\lambda_l$ is determined by the optimization method (\texttt{fanc}), given a choice for the length of the sequence $l\in \{1,2,\dots\}$.
	
	\item[S.2] Summary plot: From the DSSFA estimates obtained in step one, generate the sequence 
	\begin{equation}\label{eq_3}
	{\cal L}_{\tilde{k},\lambda}(\boldsymbol{\Omega},\boldsymbol{\hat{\Omega}}_{\tilde{k},\lambda}) = \log|\boldsymbol{\hat{\Omega}}_{\tilde{k},\lambda} |+\tr\left(\boldsymbol{\hat{\Omega}}_{\tilde{k},\lambda}^{-1}\boldsymbol{\Omega}\right),
	\end{equation}
	for $\tilde{k} = 1,\dots,k$, and $\lambda  =  \lambda_0,\lambda_1,\dots,\lambda_l$. The posterior distribution of $\boldsymbol{\Omega}$ in \eqref{eq_3} is approximated by the posterior samples $\boldsymbol{\Omega}_{(m)}$, with $\boldsymbol{\Omega}_{(m)} = \boldsymbol{B}_{(m)}\boldsymbol{B}_{(m)}^T+\boldsymbol{\Sigma}_{(m)}$, for $m = 1,2,\dots,M$. Select a quantile for the loss function of the full model, and plot the expected values $E_{\boldsymbol{\theta}|\boldsymbol{y}}[{\cal L}_{\tilde{k},\lambda}(\boldsymbol{\Omega},\boldsymbol{\hat{\Omega}}_{\tilde{k},\lambda})]$ of \eqref{eq_3}, in relation to   $\tilde{k}$ and $\lambda$ in a graphical summary.
\end{enumerate}

Ultimately, it is left to the end user to decide the best lower dimension representation given the choice of the quantile of the loss function of the full model.  Otherwise, one can use the proposed criterion of the highest expected loss generated by the highest value of $\lambda$ and lowest dimension $\tilde{k}$ that is within the quantile of the loss of the full model, resulting in an automated process. The expected value of the proposed loss, $E_{\boldsymbol{\theta}|\boldsymbol{y}}[{\cal L}_{\tilde{k},\lambda}(\boldsymbol{\Omega},\boldsymbol{\hat{\Omega}}_{\tilde{k},\lambda})]$, can be approximated by replacing the posterior parameter $\boldsymbol{\Omega}$, by $\overline{\boldsymbol{\Omega}}$ in \eqref{eq_3}. 

\subsubsection{Toy example}\label{toyexemp}

In this section, we present a toy example to illustrate our a approach and the workings of the DSSFA posterior summary. We applied our method to simulated data from a factor analysis model with loadings extracted from \cite{harman1976modern}. Originally, the loadings $\boldsymbol{B}_0$ (see Supplementary Material \ref{A2.1}) come from the analysis of eight physical variables from 305 individuals, where the true number of factors is $k_0 = 2$. In this example, we generated $n = 100$ samples from model \eqref{eq2}, with $\boldsymbol{B}_0$ and the uniqueness generated as $\boldsymbol{\Sigma}_0 = \diag(\boldsymbol{I}_{p}-\boldsymbol{B}_0\boldsymbol{B}_{0}^T)$, where $\diag(A)$ is the matrix formed by the diagonal elements of the matrix $A$. We fitted the model with independent normal priors for the loadings, and inverse gamma for the uniqueness, and set the number of factors as $k = 5$. We ran the Gibbs sampler for 10,000 iterations discarding the first 5,000 as burn-in.  Details on prior specifications are in Supplementary Material \ref{A2.2}. We obtained the posterior mean of the covariance matrix $\boldsymbol{\overline{\Omega}}$ from the posterior distribution and followed the steps presented in Subsection \ref{resum}. In the optimization, we let  $\tilde{k} = 1,2,\dots,k$, and with penalized solution path of size $\lambda  =  \lambda_0,\lambda_1,\dots,\lambda_{10}$, for each factor dimension, $\tilde{k}$. 
\begin{figure}[t!]
	\centering
	\includegraphics[width=\columnwidth]{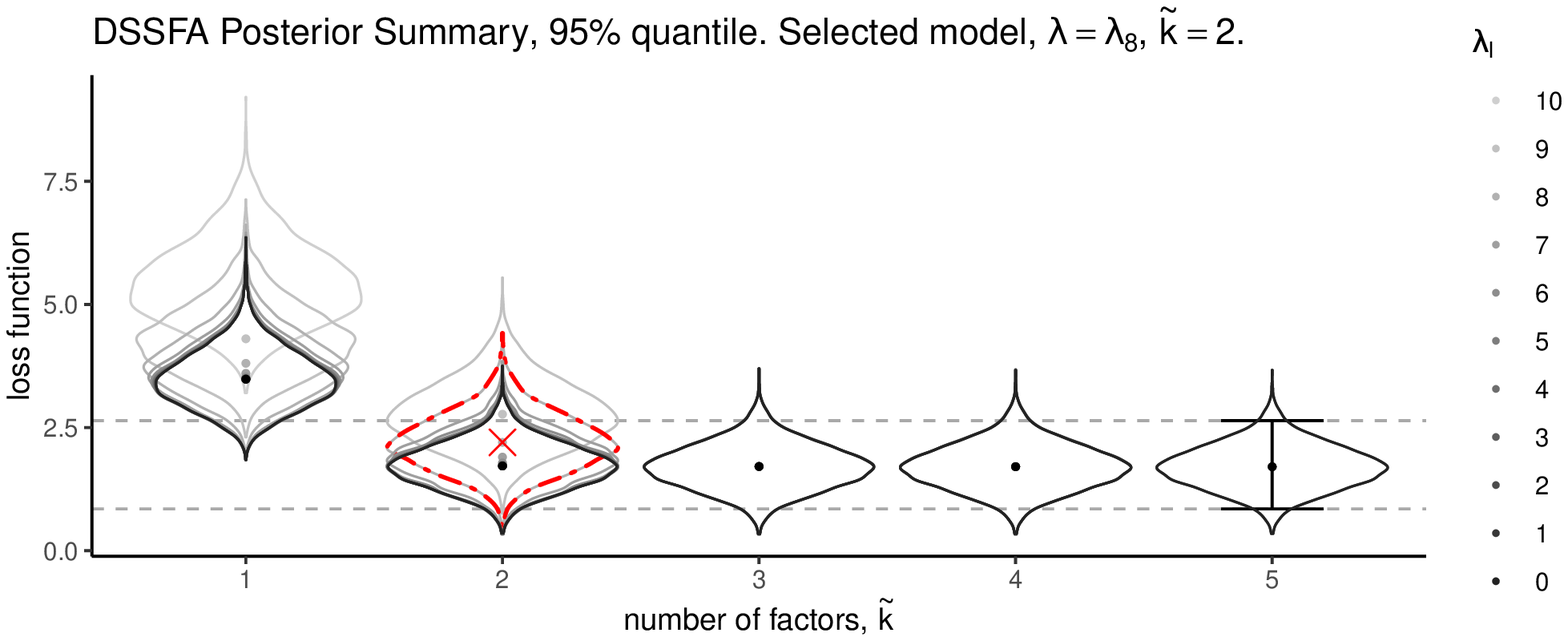}
	\caption{DSSFA summary plot for the toy example, comparing the densities of the loss functions, ${\cal L}_{\tilde{k},\lambda}(\boldsymbol{\Omega},\boldsymbol{\hat{\Omega}}_{\tilde{k},\lambda})$, defined in \eqref{eq_3} (violin plots), with respective posterior means $E_{\boldsymbol{\theta}|\boldsymbol{y}}[{\cal L}_{\tilde{k},\lambda}(\boldsymbol{\Omega},\boldsymbol{\hat{\Omega}}_{\tilde{k},\lambda})]$, (dots), indexed by the number of factors $\tilde{k} = 1,2,\dots,5$, and the penalty parameters $\lambda = \lambda_{0},\lambda_{1},\dots,\lambda_{10}$, obtained according to steps S.1 and S.2 in the method's overview in Subsection \ref{resum}. The dashed line is the 95\% quantile of the loss function of the full model  with no penalty, $\lambda_{0}=0$, identified with an error bar in $\tilde{k} = 5$. From the criterion presented in Subsection \ref{resum}, we consider the model that generates the loss function (color dot-dashed density), with the greatest expected value (identified as $\times$), that is within the 95\% quantile of the loss function of the full model. The resulted loss represents a model with $\tilde{k} = 2$ factors, and complexity parameter $\lambda  = \lambda_{8}$, which results in a loadings matrix with 19\% zeroed entries.}
	\label{fig.toy}
\end{figure}

Figure \ref{fig.toy} displays the DSSFA posterior summary plot, where the 95\% quantile of the loss of the full model is shown (dashed line). We can observe the deterioration in utility represented by the increasing values of the loss functions ${\cal L}_{\tilde{k},\lambda}(\boldsymbol{\Omega},\boldsymbol{\hat{\Omega}}_{\tilde{k},\lambda})$, in relation to $\lambda$ and $\tilde{k}$. From the posterior summary, models with $\tilde{k} = 1$ factors were not considered, since the expected posterior losses, $E_{\boldsymbol{\theta}|\boldsymbol{y}}[{\cal L}_{\tilde{k},\lambda}(\boldsymbol{\Omega},\boldsymbol{\hat{\Omega}}_{\tilde{k},\lambda})]$, were not within the 95\% quantile of the loss function of the full model. Models with $\tilde{k} \geq 2$ or greater, were considered since the values of the expected posterior losses are inside the specified quantile. Furthermore, we observe a smaller solution path generated by $\lambda$ for models with $\tilde{k}\geq 3$. This behavior arises from the fact that even for small values of $\lambda$ the optimization procedure returns loadings matrices $\boldsymbol{\hat{B}}_{\tilde{k},\lambda}$ with zeroed columns, which are discarded from the summary. The highlighted model in Figure \ref{fig.toy} is selected based on the criterion defined in Subsection \ref{gsum}, returning the complexity coefficient $\lambda = \lambda_{8}$ and the factor dimension $\tilde{k} = 2$. The selected model resulted in a loadings matrix with 19\% of its entries equal to zero. The recovered loadings matrices $\boldsymbol{\hat{B}}_{2,\lambda_{8}}$, and the loadings with no penalty, $\boldsymbol{\hat{B}}_{2,\lambda_{0}}$, can be seen in the Supplementary Material \ref{A2.3}.

\section{Simulation study}\label{sim.study}

The following study evaluates the proposed method's performance in the recovery of the true covariance matrix $\boldsymbol{\Omega}_{0} =\boldsymbol{B}_{0}\boldsymbol{B}_{0}^T+\boldsymbol{\Sigma}_{0}$, where $\boldsymbol{B}_0$ is a loadings matrix with dimension $p\times k_0$, and  $\boldsymbol{\Sigma}_{0}$ is a uniqueness matrix with dimension $p\times p$, in relation to the selection of the true number of factors $k_{0}$. We compare our proposed method with the marginal likelihood estimate for selecting the number of factors  \citep{lopes2004bayesian,newton1994approximate}, and Bayes model averaging (BMA) \citep{hoeting1999bayesian} for covariance matrix estimation. We provide evidence that DSSFA improves the selection of the number of factors, in the same posterior samples, without affecting the estimation of $\boldsymbol{\Omega}_0$, in data generated from normal, and non-normal factor models.

\subsection{Simulation settings}\label{sim_settings}

We followed the simulation design of \cite{man2020mode}, in which the synthetic data $\boldsymbol{y} = (\boldsymbol{y}_1,\dots,\boldsymbol{y}_n)^T$ are drawn from model \eqref{eq2}, with sample sizes $n = 100, 500,1000$, with $p = 15$ variables and $k_0 = 3$ factors.  The entries of the loadings matrix $\boldsymbol{B}_0$ were independently drawn from a standard normal distribution $N(0,1)$ for each replicate. The resulting loadings were rotated to be positive lower triangular (PLT) to assure identifiability. The idiosyncratic matrix is set as $\boldsymbol{\Sigma}_0 = \sigma^2\boldsymbol{I}_{p}$, with $\sigma = 0.2,0.5$. We considered two setups for the latent factors and idiosyncratic errors. In the first setup, we investigated the standard factor analysis model with $\boldsymbol{f}_i\sim N_{k_0}(\boldsymbol{0},\boldsymbol{I}_{k_0})$, and $\boldsymbol{\epsilon}_j\sim N_p(\boldsymbol{0},\boldsymbol{\Sigma}_0)$, for $i = 1,2,\dots,n$, and $j = 1,2,\dots,p$. In the second setup, we explored the robustness of our approach by using the standard multivariate $t$-distribution, $t_{k_0}(0,\boldsymbol{I}_{k_0},\nu)$ for $\boldsymbol{f}_i$, and the scaled $t$-distribution $t_p(\boldsymbol{0},\boldsymbol{\Sigma}_0,\nu)$ for $\boldsymbol{\epsilon}_j$, with $\nu = 3,10$, degrees of freedom. Under this setup, we evaluated our method on data of size $n = 100$. See Supplementary Material \ref{A3.2} for simulation results from the non-normal data. 

We included three different priors in the study. The first is from \cite{geweke1996measuring}, which uses the standard PLT constraint on the loadings. The second, is a novel prior presented by \cite{man2020mode}, which incorporates a PLT type constraint with a mode-jumping step to avoid multimodal posteriors. The third is the plain normal prior on the loadings without constraints. We ran Monte Carlo Markov Chain (MCMC) algorithms for 30,000 iterations discarding the first 15,000 as burn-in. For each algorithm we generate posteriors with factor dimensions $k = 1,2,\dots,5$. To generate the posterior samples from the chosen priors, we used algorithms from the supplementary material of \cite{man2020mode}. For details on parameter and hyper-parameter specifications see Supplementary Material \ref{A3.1}.  We ran 300 replicates for each simulation scenario, under the normal and non-normal errors.

\subsection{Evaluation and results}

In this section, we followed the steps presented in Subsection \ref{resum}, and applied the DSSFA method only on the posterior samples with factor dimension $k = 5$. Hence, we considered $\tilde{k} = 1,2,\dots,5$ factors, and  set $\lambda = 0$, in the optimization procedure since we are not interested in the effects of sparse loadings in this study. We used the criterion described in Subsection \ref{resum}, and let the method auto-select point estimates under 95\% and 99\% quantiles of the loss of the full model. As a result, we obtained the factor dimension $\tilde{k}$ and the DSSFA estimates  $\boldsymbol{\hat{B}}_{\tilde{k}}$ and $\boldsymbol{\hat{\Sigma}}_{\tilde{k}}$, from which we recover the covariance matrix estimate, $\boldsymbol{\hat{\Omega}}_{\tilde{k}} = \boldsymbol{\hat{B}}_{\tilde{k}}\boldsymbol{\hat{B}}_{\tilde{k}}^T+\boldsymbol{\hat{\Sigma}}_{\tilde{k}}$. Conversely to our procedure, we use bridge sampling to calculate the marginal likelihood estimate for each posterior with dimensions $k = 1,2,\dots,5$. We recall that bridge sampling is recommended by \cite{lopes2004bayesian} and \cite{man2020mode}, over other methods, including the harmonic mean and Newton-Raftery's estimators \citep{newton1994approximate}.  From the same marginal likelihood estimates we obtained the weights and recovered the BMA estimates of the covariance matrix. Finally, we used the root mean squared error (RMSE) defined as $\text{RMSE} = (1/p^2 \sum_{j=1}^p\sum_{q=1}^p(\boldsymbol{\bar{\Omega}}_{jq}-\boldsymbol{\Omega}_{0jq})^2)^{1/2}$ to evaluate the recovery of the true covariance matrix $\boldsymbol{\Omega}_0$, where $\boldsymbol{\bar{\Omega}}$ is the recovered point estimates from the two procedures.
\begin{table}[t!]
	\footnotesize
	\caption{Proportions of correctly identified models, in percentage, for 300 simulation replicates from a normal factor model with standard deviations $\sigma = 0.2, 0.5$, with sample sizes $n = 100, 500,1000$, and true factor dimension $k_0 = 3$.  The number of factors was identified using the DSSFA method, with 95\% and 99\% quantiles of the loss function of the full model, and the marginal likelihood (MargLike) using bridge sampling. We considered the Geweke \& Zhou, Unconstrained and Man \& Culpepper priors in the simulation.}
	\centering
	\label{tab1}
	\begin{tabular}{lllccccccccc}
		&   &  & &\multicolumn{8}{c}{Prop. of correctly identified models, (\%)}\\
		\cline{5-12} 
		&   &  & &\multicolumn{2}{c}{$n=100$} & &  \multicolumn{2}{c}{$n = 500$} & &  \multicolumn{2}{c}{$n = 1000$}\\ 
		\cline{5-6} \cline{8-9}   \cline{11-12}  
		Standard deviation $\sigma$  & & & & 0.2 & 0.5 & & 0.2  & 0.5 && 0.2 & 0.5 \\
		
		\hline
		prior    &  method    &  quantile   & &  &  &   &  &   &  & &\\
		
		Geweke \& Zhou    &  DSSFA    & 95\%   & &100 & 100 & & 100& 100 & & 100& 100\\
		&                & 99\%   & &  100& 100 & & 100& 100 & & 100& 100 \\
		& MargLike  &           & &  71& 80 & & 78& 86 & & 84& 84 \\
		
		Unconstrained  &  DSSFA   & 95\% & & 100 & 100 & & 100& 100 & & 100& 100\\
		&                & 99\% & &  100& 100 & & 100& 100 & & 100& 100\\
		& MargLike &           & &  100& 100 & & 100& 100 & & 100& 100\\
		\
		Man \& Culpepper  &  DSSFA    & 95\% & & 100 & 100 & & 100& 100 & & 100& 100 \\
		&                & 99\% & & 100& 100 & & 100& 100 & & 100& 100 \\
		& MargLike  &          & &  100& 100 & & 100& 100 & & 100& 100 \\
		\hline		     
	\end{tabular}
\end{table}	

Table \ref{tab1} shows the proportions, in percentage, of correctly identified models by the methods. The marginal likelihood estimate using bridge sampling selected the incorrect number of factors under the \cite{geweke1996measuring} prior in all scenarios. \cite{man2020mode} argue that some of the uncertainty in the marginal likelihood estimation may be related to the posterior multi-modality generated by the standard PLT constraint. The DSSFA method circumvents this problem by selecting 300 out of 300 replicates in both credible intervals and all simulation scenarios under this prior. Furthermore, for samples from \cite{man2020mode} and the Unconstrained priors, our method selected the correct number of factors in every single replicate in both credible intervals and all setups. This result was repeated under marginal likelihood estimation. 

\begin{figure}[t!]
	\centering
	\includegraphics[width=\columnwidth]{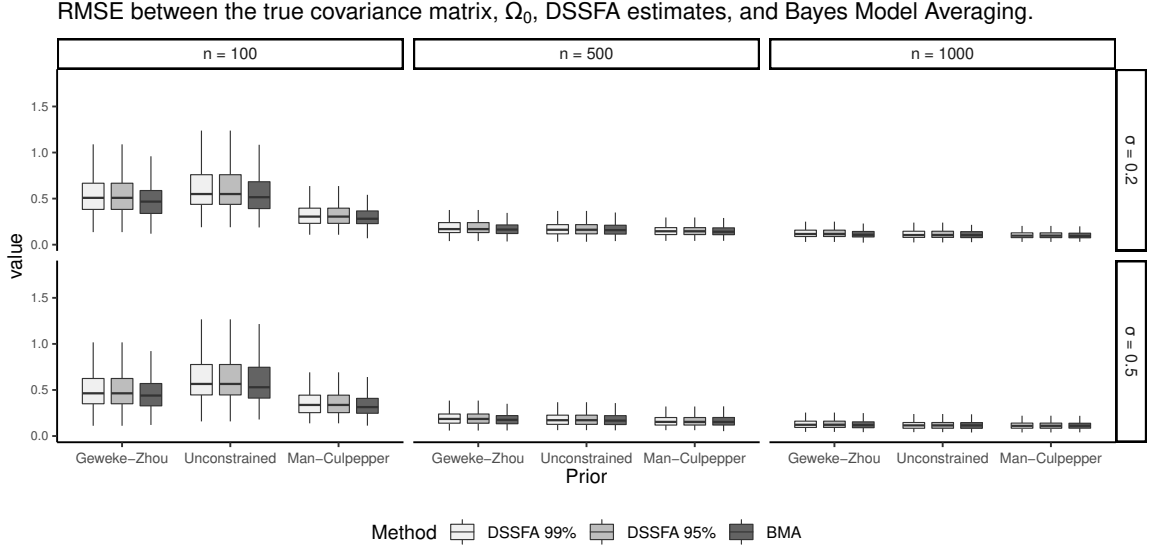}
	\caption{Boxplots of RMSE between the true covariance matrix $\boldsymbol{\Omega}_0$, and covariance estimates generated by the Methods: DSSFA, with 95\% and 99\% quantiles of the loss function of the full model, and Bayes model averaging (BMA), with priors, Geweke \& Zhou, Unconstrained  and Man \& Culpepper. The 300 replicates were generated from the normal factor analysis model, with factor dimension $k_0 = 3$, uniqueness $\sigma = 0.2,0.5$, and sample sizes $n = 100,500,1000$.}
	\label{fig3}
\end{figure}

Figure \ref{fig3} displays the RMSE of the covariance matrix recovery for the same 300 replicates. We observe no significant loss in fit between the DSSFA estimates in relation to the weighted posterior mean obtained using BMA. Moreover, we can observe that the DSSFA estimates conserve characteristics of the posterior distributions. The DSSFA estimates recovered under the \cite{man2020mode} prior outperforms the one in \cite{geweke1996measuring} and the Unconstrained in fit, confirming the results from \cite{man2020mode}. Lastly, we observe that our method successfully estimates $\boldsymbol{\Omega}_0$ as the sample size increases.

\section{Real data analysis}\label{data.ana}

This section aims to show the flexibility and usefulness of our method through an application to personality traits data. We obtained an estimate for the number of factors by applying the DSSFA method in posterior samples generated from over-fitted priors for factor analysis. These priors starts with a conservative factor dimension, and remove components by shrinking their loadings to zero, and select the number of factors by using adaptive Gibbs sampling methods. Such approaches included in this category are the multiplicative gamma process (MGP) from \cite{bhattacharya2011sparse}, and a novel procedure that uses cumulative shrinkage priors (CUSP), from  \cite{legramanti2020bayesian}. As seen in Subsection \ref{expec_dssfa}, we can use our approach in such models since the DSSFA method is agnostic to prior choice, and depends uniquely on the posterior distribution of the covariance matrix. For the data analysis, we followed \cite{legramanti2020bayesian} and explored a subset of the big five personality traits data, which is available at \texttt{bfi} in the \texttt{R} package \texttt{psych}. We examine the association structure among $p = 25$ personality variables collected from $n = 126$ individuals over age 50. We also centered the data and changed the sign of the variables 1, 9, 10, 11, 12, 22 and  25. Additional details on the prior specification of the two models are reported in the Supplementary Material \ref{A4.1}. The number of factors were initialized as $k = p$ for MGP, and as $k = p+1$ for CUSP. We ran the MCMC algorithms for 10,000 iterations discarding the first 5,000 as burn-in and with thinning in every five samples. At a first run, MGP obtained a posterior mean (95\% credible interval) of 20.7 ([18,24]) for the number of factors, while CUSP obtained 2.64 ([2,3]).
\begin{figure}[t!]
	\centering
	\includegraphics[width=\columnwidth]{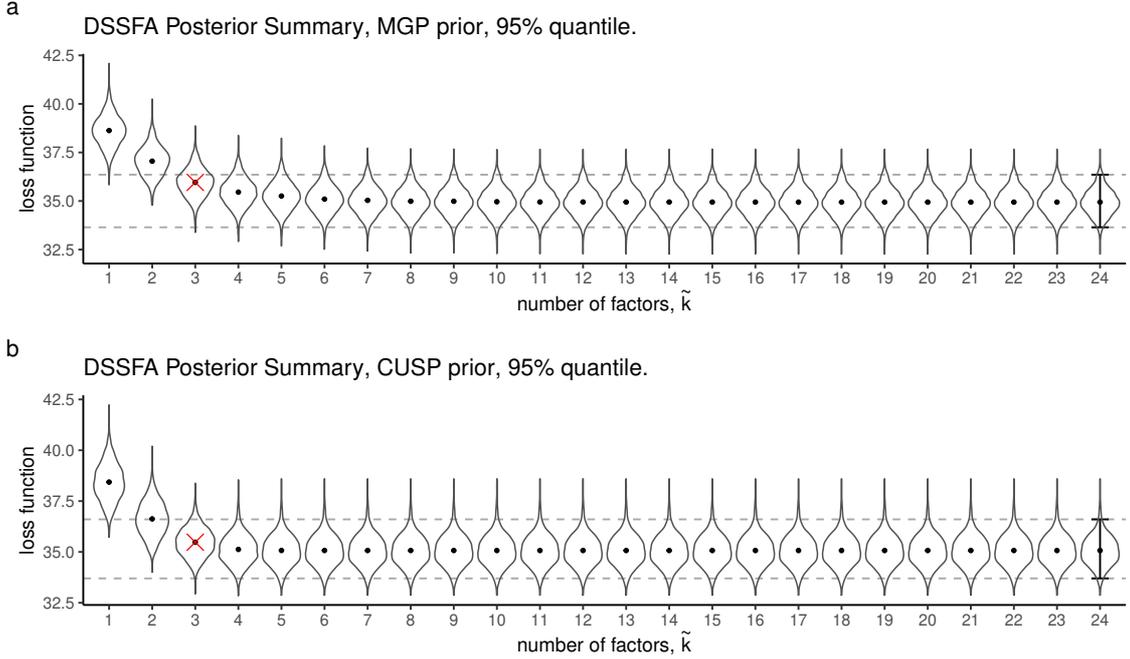}
	\caption{DSSFA posterior summary plots generated with (a) MGP, and (b) CUSP priors, obtained according to the steps presented in Subsection \ref{resum}. The plots display the densities of the loss functions ${\cal L}_{\tilde{k},\lambda}(\boldsymbol{\Omega},\boldsymbol{\hat{\Omega}}_{\tilde{k},\lambda})$ (violin plots), with respective posterior means $E_{\boldsymbol{\theta}|\boldsymbol{y}}[{\cal L}_{\tilde{k},\lambda}(\boldsymbol{\Omega},\boldsymbol{\hat{\Omega}}_{\tilde{k},\lambda})]$ (dots), indexed by the number of factors $\tilde{k} = 1,2,\dots,24$. No penalty was used. The dashed line is the 95\% quantile of the loss  function of the full model generated with $\tilde{k} = 24$ (error bar).  Following the criterion in Subsection \ref{resum}, we consider the model that generates a loss function whose expected value, (identified as $\times$), is within the selected quantile of the loss function of the full model. The resulted loss function represents a model with $\tilde{k} = 3$ factors, for the MGP, and for CUSP priors.}
	\label{fig.app1}
\end{figure}

From the same posterior samples generated by MGP and CUSP, we followed the steps presented in Subsection \ref{resum}, and applied the DSSFA method with $\tilde{k} = 1,2,\dots, (p-1)$ factors for the two priors. This limit is imposed by the used optimization procedure (\texttt{fanc}). We let our method auto-select the factor dimension under the 95\% quantile of the loss function of the full model, and no regularization was used, $\lambda = 0$. Figure \ref{fig.app1} displays the posterior summary plots for the two methods.  Under a 95\% quantile of the loss function of the  full model, our method selected a factor model of size three for MGP and CUSP. These results are in agreement with the analysis of \cite{legramanti2020bayesian}, in which three main factors were identified. Further analysis, on the posterior summary plots indicate that the MGP posterior contains information of a three-factor model, although the posterior adaptation procedure privileges a model with redundant factors.

\begin{figure}[t!]
	\centering
	\includegraphics[width=\columnwidth]{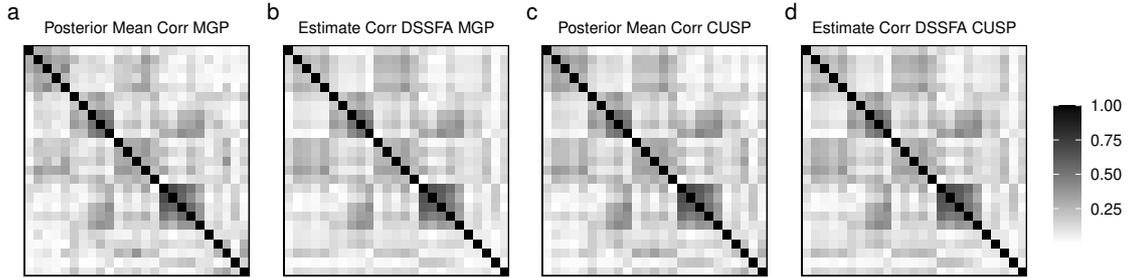}
	\caption{Absolute values of the posterior mean of the correlation matrix under the (a) MGP, and (c) CUSP priors, compared with the absolute correlation matrix estimates produced by the DSSFA method with 95\% quantile of the loss function of the full model, under (b) MGP, and (d) CUSP priors.}
	\label{fig.app3}
\end{figure}

Figure \ref{fig.app3} displays the absolute values of the posterior mean of the correlation matrix, from priors MGP, (a), and CUSP, (c), in comparison to the absolute values of the correlation matrix generated by the DSSFA estimates with the 95\% quantile of the loss function of the full model under (b) MGP and (d) CUSP. We can observe that the values of the absolute correlations are similar across the different estimates, which indicates that MGP overestimates the number of factors, further confirming the analysis of \cite{legramanti2020bayesian}, and the results from the DSSFA posterior summary plot in Figure \ref{fig.app1}.

\begin{figure}[t!]
	\centering
	\includegraphics[scale=0.5]{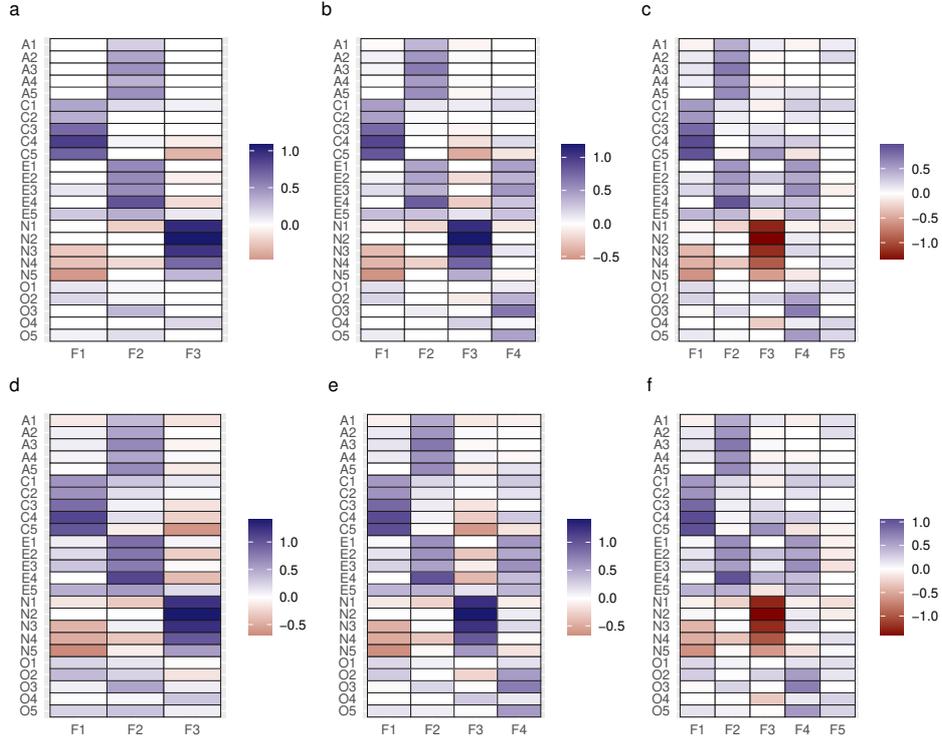}
	\caption{DSSFA estimates under CUSP prior for the personality traits data. Figures (a) and (d) show the resulting sparse loadings $\boldsymbol{\hat{B}}_{3,\lambda_8}$, and loadings without penalty $\boldsymbol{\hat{B}}_{3,\lambda_0}$ selected with the DSSFA procedure, respectively. We also include in our analysis models with $\tilde{k} = 4$ factors with penalty, $\boldsymbol{\hat{B}}_{4,\lambda_7}$ in (b), and without penalty, $\boldsymbol{\hat{B}}_{4,\lambda_0}$ in (e), and models with $\tilde{k} = 5$ factors with penalty $\boldsymbol{\hat{B}}_{5,\lambda_5}$ in (c), and without penalty $\boldsymbol{\hat{B}}_{5,\lambda_0}$ in (f). The presented models are within the 95\% quantile interval of the loss function of the full model (Figure \ref{fig_supp} in Supplementary Material \ref{A4.2}). The percentage of zeroed loadings in the regularized models in (a), (b), and (c) are 39\%, 27\%, and 20\%, respectively.}
	\label{fig.ap2}
\end{figure}

In the factor analysis literature, over-fitted factor models are usually used for covariance matrix estimation, and thus there is no need to focus on the identifiability and interpretation of the resulting factor loadings structure. We went a step further in our analysis, and used DSSFA with penalty on the loadings, and same posterior samples generated by the CUSP prior in the previous study, in order to recover an interpretable loadings matrix. We use the penalty on the loadings in order to prevent the non-identifiability of the loadings \citep[see][]{scharf2019should}. We run our method with $\tilde{k} = 1,2,\dots, (p-1)$ factors, with a penalized solution path of size $\lambda = \lambda_0,\lambda_1,\dots,\lambda_{10}$, for each factor dimension, $\tilde{k}$. The DSSFA posterior summary plot can be seen in Figure \ref{fig_supp}, in the Supplementary Material \ref{A4.2}. Under a 95\% quantile of the loss of the full model, our method selected a factor analysis model with dimension $\tilde{k} = 3$. Models with $\tilde{k}\leq 2$ were not considered. We also included in our analysis models with $\tilde{k} = 4$ and $\tilde{k} = 5$ factors, since they are within the 95\% quantile of the loss function of the full model. Figure \ref{fig.ap2} displays the selected loadings, in which we can observe a similar factor structure across the different dimensions, although some of the entries of the penalized models are shrunken in comparison with the entries from the models with no penalty ($\lambda_{0} = 0$). 

As in \cite{legramanti2020bayesian}, we notice significant correlation between agreeableness (A) and extraversion (E) in factor $F_2$, some negative correlation between conscientiousness (C) and neuroticism (N) in factor $F_1$. Lastly, Openness (O) presented less evident weights and almost no association between traits. Furthermore, the small values of the loadings in Openness may be associated with the chosen subset of the data (age $>$ 50), suggesting that this trait is less present in the considered age group.

\section{Conclusions}\label{concl}

In this paper, we introduced the DSSFA method, which advances the factor analysis literature by introducing an approach that specifies posterior summarization as a decision 	problem. The proposed method has two steps: optimization of a predefined loss function and  a posterior summary plot. Unlike hard thresholding rules and classical methods based on information criteria, our method brings the possibility of simultaneously selecting the model and factor loadings.  Furthermore, our procedure is extremely flexible, being used in conjunction with whichever prior distribution is most appropriate to the problem on the condition that posterior samples of the Bayesian factor analysis model are available. 

The posterior summary plots revealed the relation between posterior uncertainty, sparsity and predictive degradation. From these relations, we proposed a criterion that automates the problem of determining the number of factors and the complexity penalty. We performed Monte Carlo experiments based on the this criterion that provided evidence of improvement model selection over other widely used methods, also extended to data generated from non-normal factor models. The usefulness of our model was further assessed by exposing redundant factors in over-fitted factor models in the application. Moreover, to the best of our knowledge, this is the first time interpretable loadings matrices are extracted from models in which the posterior samples have different factor dimensions.  However, one downside of our approach is the over shrinkage of the loadings estimates of the recovered sparse models.

Future research should consider other different loss functions than the negative log-likelihood. Viable options include squared loss, and the Frobenius distance. Another criterion for model selection can also be easily be explored. One possibility, in situations with a considerable size of variables, such as in genetic data, is that the optimization procedure can run increasingly and stop as soon as the expected loss is within the quantile of the loss of the full model. A natural extension of this criterion would be an application to high-dimensional data. Other penalties could be used to circumvent the problem of over shrinkage of the loadings observed in the toy example and application. A possible alternative is the Bayesian adaptive penalty presented recently by \cite{kowal2020bayesian}.  Finally, we envisage the extension of the DSSFA approach to other latent variable models \citep{bartholomew2011latent}.

%\vspace{0.3cm}

\section*{Supplementary Material}

Supplementary material includes implementation the optimization method, further simulation results and prior specifications. Code and data used in this paper are available at \texttt{https://github.com/hbolfarine/dssfa}.

\section*{Funding}

Henrique Bolfarine gratefully acknowledge support from CAPES (Coordena\c c\~ao de Aperfei\c c oamento de Pessoal de N\'ivel Superior), grant number 88887.571312/2020-00 and from the Salem Center for Policy at the
University of Texas at Austin McCombs School of Business. Hedibert F. Lopess thank FAPESP (Funda\c c\~ ao de Amparo \` a Pesquisa do Estado de S\~ ao Paulo) for financial support through grant number 2017/10096-6.

\bibliographystyle{plainnat}

\appendix

%\numberwithin{equation}{section}
 \renewcommand{\theequation}{\thesection.\arabic{equation}}
 
  \renewcommand{\thepage}{S\arabic{page}}
 \renewcommand{\thesection}{S\arabic{section}}
 \renewcommand{\thetable}{S\arabic{table}}
 \renewcommand{\thefigure}{S\arabic{figure}}

\section*{Supplementary Material}

\section{Optimization method}\label{A1}

\subsection{Implementation of the DSSFA method}\label{A1.3}

To perform the optimization, we first replace the sample covariance by $\overline{\boldsymbol{\Omega}}$ in the package's main function (\texttt{fanc}). The  function uses as a default the MC+ penalty \citep{zhang2010nearly}, which is a non-convex function indexed by $\texttt{rho}>0$.  To obtain the soft threshold, let $\texttt{rho} \rightarrow \infty$  in the arguments of the function. 

\section{Toy example details}\label{A2}

\subsection{Loadings matrix}\label{A2.1}
\begin{equation*}
\boldsymbol{B}_{0} =
\left(
\begin{array}{cccccccc}
0.879&0.919&0.890&0.858&0.238&0.183&0.135&0.250\\
0.272&0.210&0.182&0.246&0.900&0.792& 0.729&0.684
\end{array}
\right)^T,
\end{equation*}

\subsection{Prior Specification}\label{A2.2}

The normal distribution $N(0,\eta)$ was assigned to the loadings $\boldsymbol{B}_{0}$, with $\eta$ and $\sigma_j$ following an Inverse Gamma distribution ($\eta,\sigma_j\sim IG(1,1)$), for $j = 1,2,\dots,p$, for the idiosyncratic variances. The Gibbs sampler for the factor analysis algorithm is implemented in Rcpp, and is available in the supplementary material of \cite{man2020mode}.

\subsection{Estimated Loadings}\label{A2.3}

\begin{figure}[ht]
	\centering
	\includegraphics[scale=0.5]{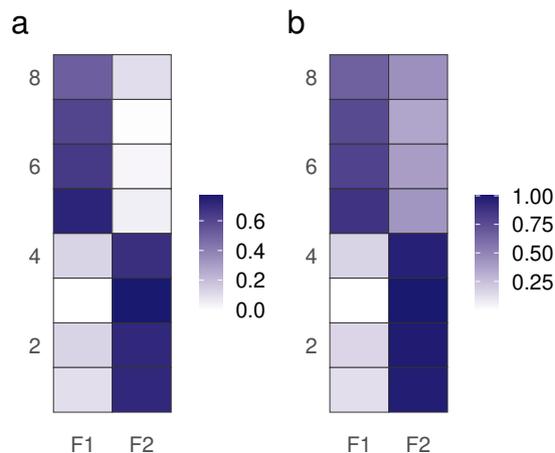}
	\caption{DSSFA estimates of the loadings matrix with sparse representation (left), and without regularization (right).}
	\label{fig.app2}
\end{figure}

\section{Additional simulation and details}\label{A3}

\subsection{MCMC model specifications}\label{A3.1}

We followed the prior and hyperparameter specifications used in the study of \cite{man2020mode}. In the algorithm from \cite{geweke1996measuring} the loadings were assigned independent normal priors $b_{jk}|\eta \sim N(0,\eta)$, for $j = 1,\dots,p$, and $k$ the fixed number of factors, and are restricted as PLT, where $b_{jk} = 0$ for $k>j$, and $b_{kk}>0$. \cite{man2020mode} relaxed the standard constraint of \cite{geweke1996measuring}, and applied the PLT constraint 
to any arbitrary subset of the $p$ rows of the loadings matrix. Hence, for a permutation set $\boldsymbol{r} = (r_1,r_2,\dots,r_{k})\subset \{1,2,\dots,p\}$, with $r_k \neq r_{k'}$, and $k\neq k'$, the loadings were constrained as $b_{r_{k}k'} = 0$ for $k'>k$, and $b_{r_{k}k}>0$. To avoid near singular cases in the sub matrix generated by  $\boldsymbol{r}$, \cite{man2020mode} used the prior distribution, $p(b_{r_{k}k}|\eta,\gamma) \propto b_{r_{k}k}^{\gamma}\exp(-b_{r_{k}k}^2 /2\eta)$, for $b_{r_{k}k}>0$, where $\gamma = 0.5$. A sample from the non-singular restricted PLT submatrix is obtained with a  Metropolis-Hastings step, which is incorporated into the loadings matrix via matrix decomposition \citep[see][]{man2020mode}. Lastly, an uniform prior was assigned to the permutation set $\boldsymbol{r}$. The normal distribution $N(0,\eta)$ was assigned to the remaining unrestricted loadings. In the Unrestricted model, the entirety of loadings was assigned the normal distribution $N(0,\eta)$ prior. In all models, $\eta \sim IG(1,1)$ was chosen for the variance of the loadings, and $\sigma_{j}\sim IG(1,1)$, for $j = 1,2,\dots,p$, for the idiosyncratic variances. The MCMC implementations are in Rcpp, and are available in the supplementary material of \cite{man2020mode}.

\subsection{Simulation Results - non-normal data}\label{A3.2}

Table \ref{tab2} shows the percentage of the number of times the models with dimension $k = 1,2,\dots,5$, were selected in each simulation replicate by the chosen methods for data generated from the non-normal factor analysis model. The true number of factors was fixed as $k_0 = 3$. For data generated with $\nu  = 3$, our proposed method outperformed the marginal likelihood estimate in all  simulation scenarios and priors. This difference is more apparent when the 99\% quantile of the full model is used to select the DSSFA estimates. This result may come from the fact that the resulted loss function reflects the noise contained in the data, and thus being more dispersed and making a larger quantile more effective. 

\begin{table}[ht]
	\footnotesize
	\caption{ Proportion of correctly identified models, in percentage, for 300 simulations of a normal factor model with standard deviations $\sigma = 0.2, 0.5$, with sample size $n = 100$, and true factor dimension $k_0 = 3$.  The number of factors were identified using the Methods: DSSFA, with 95\% and 99\% quantiles, and marginal likelihood (MargLike) using bridge sampling. The priors considered in the simulation were the Geweke \& Zhou, Unconstrained and Man \& Culpepper.}
	\centering
	\label{tab2}
	\begin{tabular}{lllcccccc}
		&   &  & &\multicolumn{5}{c}{Proportion, (\%)}\\
		\cline{5-9} 
		&   &  & &\multicolumn{2}{c}{$\nu=3$} & &  \multicolumn{2}{c}{$\nu = 10$}\\ 
		\cline{5-6} \cline{8-9}  
		Standard deviation $\sigma$  & & & & 0.2 & 0.5 & & 0.2  & 0.5 \\
		%         \hline
		\hline
		prior    &  method    &  quantile   & &  &  &   &  &\\
		%		\hline
		Geweke \& Zhou    &  DSSFA    & 95\%   & &70 & 61 & & 100& 100\\
		&                & 99\%   & &  80& 74 & & 100& 100 \\
		& MargLike  &           & &  40& 23 & & 80& 81 \\
		%		\hline
		Unconstrained  &  DSSFA   & 95\% & & 74 & 68 & & 100& 100\\
		&                & 99\% & &  82& 80 & & 100& 100 \\
		& MargLike &           & &  52& 32 & & 100& 100\\
		%		\hline
		Man \& Culpepper  &  DSSFA    & 95\% & & 64 & 61 & & 100& 100  \\
		&                & 99\% & & 76& 72 & & 100& 100 \\
		& MargLike  &          & &  44& 30 & & 100& 100\\
		\hline		     
	\end{tabular}
\end{table}

\begin{figure}[ht]
	\centering
	\includegraphics[width=\columnwidth]{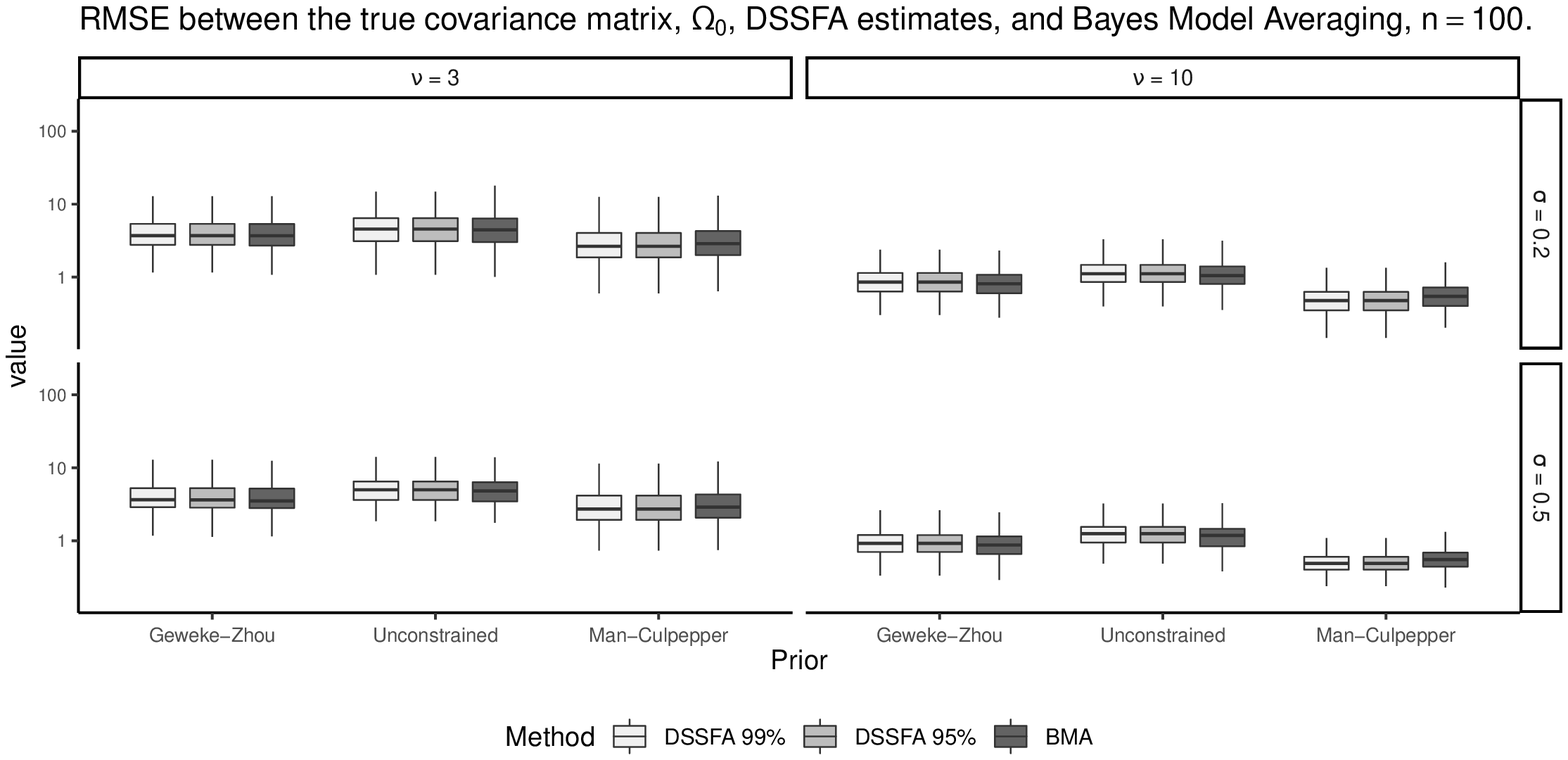}
	\caption{RMSE between the true covariance matrix $\boldsymbol{\Omega}_0$, and the estimates generated by the methods, DSSFA, with 95\% and 99\% quantiles of the loss of the full model, and Bayes model averaging (BMA), with priors: Geweke \& Zhou, Unconstrained, and Man \& Culpepper. The 300 replicates were generated from factor analysis model, with $t$-student error, with degrees of freedom $\nu = 3,10$, with factor dimension $k_0 = 3$, with uniqueness $\sigma = 0.2,0.5$, and with sample size $n = 100$. The  y-axis is on the log-scale.}
	\label{fig1}
\end{figure}

Finally, we note that as the degrees of freedom $\nu$ increase from 3 to 10 the DSSFA, and the marginal likelihood estimates have a similar performance for data generated from with normal factors and errors. Figure \ref{fig1} displays a summary of the results of the root mean squared error (RMSE) from the estimates of $\boldsymbol{\Omega}_{0}$ for 300 replicates across different scenarios. As in the results from the simulation study in Section \ref{sim.study}, we can observe that the estimates generated by DSSFA are directly related to the different posterior distributions, and thus there is no loss in the covariance matrix estimation. 

\section{Personality data details}\label{A4}

\subsection{Prior specifications}\label{A4.1}

In the multiplicative gamma process (MGP)  \citep{bhattacharya2011sparse}, the loadings were distributed as $N(0,\phi_{jk}^{-1} \theta_{k}^{-1})$, for $j = 1,\dots,p$ and  $k \in \{1,2,\dots\}$, with $\phi_{jk}\sim Gamma(3/2,3/2)$. For the global precisions $\theta_{k}^{ - 1}$ we have the multiplicative gamma process prior $\theta_{k} = \vartheta_1\cdots \vartheta_{k}$, with $\vartheta_1\sim Gamma(2.1,1)$ and $\vartheta_l \sim Gamma(3.1,1)$, for $l \geq 2$ and $k \in \{1,2,\dots\}$. The second prior is the cumulative shrinkage prior (CUSP), presented in \cite{legramanti2020bayesian}. This prior induces increasing shrinkage via a sequence of spike-and-slab distributions that assign growing mass to the spike as the model complexity grows. The  factor loadings $b_{jk}$ are distributed as $N(0,\theta_k)$ for $k \in \{1,2,\dots\}$, with $\theta_k$ assuming the spike and slab mixture $\theta_k|\pi_k \sim (1-\pi_k)IG(2,2)+\pi_k\delta_{\theta_{\infty}}$, where $\pi_k = \sum_{l = 1}^k \omega_l$, with  $\omega_l = \nu_l \prod_{m = 1}^{l-1}(1-\nu_{m})$, and $\nu_q \sim Beta(1,5)$. The spike is defined on $\delta_{\theta_{\infty}}$, which is the point mass on $\theta_{\infty}$, with $\theta_{\infty} = 0.05$. For the two methods, we have $\sigma_{j}^2 \sim IG(1,0.3)$,  for $j=1,\dots,p$, for the idiosyncratic variances.  The adaptation $p(t) = \exp(\alpha_0+\alpha_1 t)$ was allowed only after $t = 500$ iterations and $(\alpha_0, \alpha_1)$ were set to $(-1,-5\times 10^{-4})$, while the adaptation threshold $\epsilon$ in the MGP is $10^{-4}$. The MGP method was sampled using the \texttt{R} package \texttt{infinitefactor}, and CUSP was generated from the algorithm at \texttt{https://github.com/siriolegramanti/CUSP}. 

 \clearpage 
\subsection{Summary plot for personality traits data}\label{A4.2}

\begin{figure}[!h]
	\centering
	\includegraphics[width=\columnwidth]{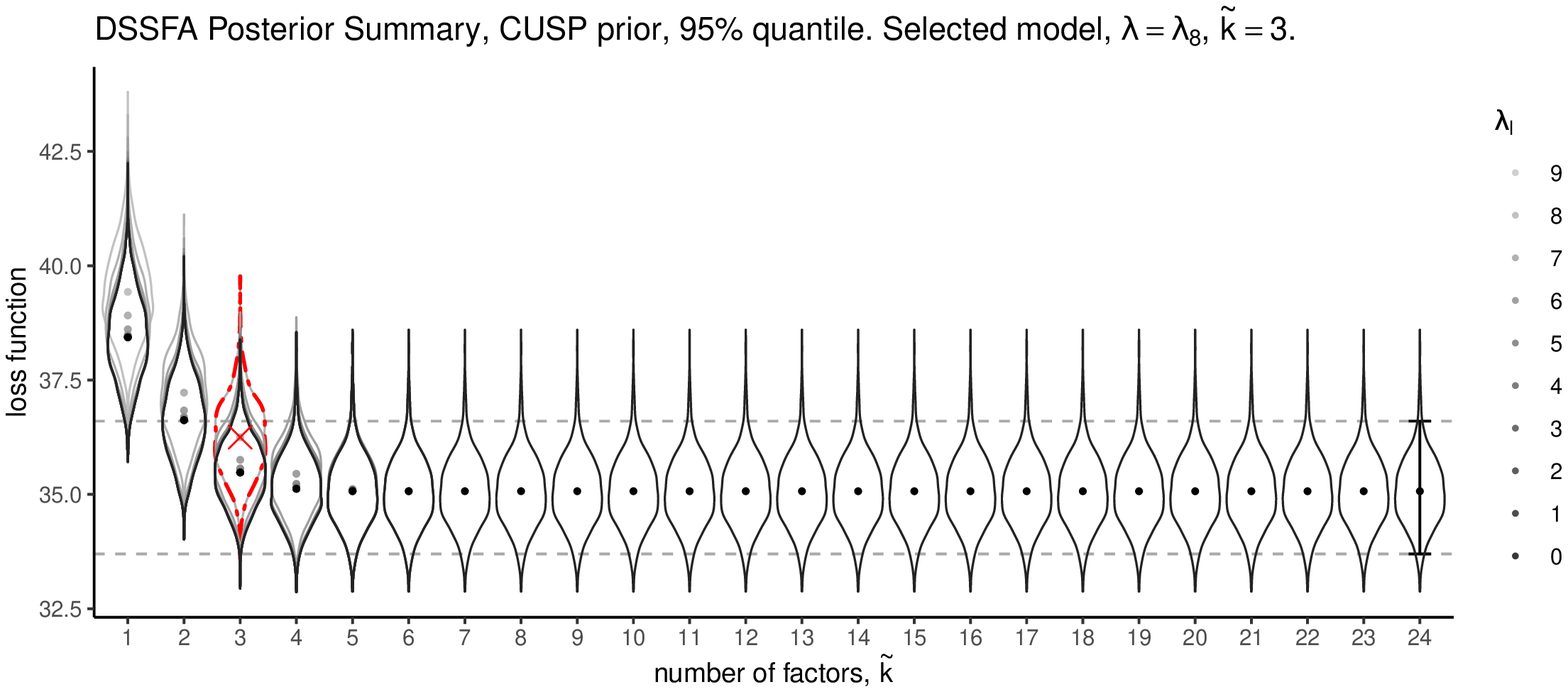}
	\caption{DSSFA summary plot for personality traits data under the CUSP prior comparing the densities of the loss functions ${\cal L}_{\tilde{k},\lambda}(\boldsymbol{\Omega},\boldsymbol{\hat{\Omega}}_{\tilde{k},\lambda})$ (violin plots), with respective posterior means $E_{\boldsymbol{\theta}|\boldsymbol{y}}[{\cal L}_{\tilde{k},\lambda}(\boldsymbol{\Omega},\boldsymbol{\hat{\Omega}}_{\tilde{k},\lambda})]$ (dots), indexed by the number of factors $\tilde{k} = 1,2,\dots,24$, and the complexity parameters $\lambda = \lambda_{0},\lambda_{1},\dots,\lambda_{10}$, obtained according to S.1 and S.2, in the Method's overview in Subsection \ref{resum}. The dashed line is the 95\% quantile of the loss function of the full model (error bar), with $\tilde{k} = 24$, and with regularization $\lambda_0=0$. Following the DSSFA method criterion presented in Subsection \ref{resum}, we consider the model that generates the loss function (color dot-dashed density), with the greatest expected value (identified as $\times$), that is within the 95\% quantile of the loss function of the full model. The resulted loss represents a model with $\tilde{k} = 3$ factors, and parsimony parameter $\lambda = \lambda_{8}$, which results in a loadings matrix with 39\% zeroed entries.}
	\label{fig_supp}
\end{figure}

\end{document}